\documentclass[twocolumn]{article}

\usepackage[resetlabels,labeled]{multibib}

\newcites{S}{References}
\usepackage{titling}
\usepackage[english]{babel}

\usepackage[letterpaper,top=2cm,bottom=2cm,left=3cm,right=3cm,marginparwidth=1.75cm]{geometry}
\usepackage{authblk}
\usepackage[T1]{fontenc}
\usepackage{siunitx}
\usepackage{cite}
\DeclareSIUnit\sq{Sq}
\DeclareSIUnit\OSq{\ohm \per \sq}
\DeclareSIUnit\Torr{Torr}
\DeclareSIUnit\mt{\milli\Torr}

\usepackage{amsmath}
\usepackage{graphicx}
\usepackage{float}
\usepackage[colorlinks=true, allcolors=blue]{hyperref}

\title{Monolithic High Contrast Grating Integrated with Metal: Infrared Electrode with Exceptionally High Conductivity and Transmission}

\author[1]{Marek Ekielski}
\author[2]{Weronika Głowadzka}
\author[1,2]{Karolina Bogdanowicz}
\author[3]{Michał Rygała}
\author[3]{Monika Mikulicz}
\author[2]{Patrycja Śpiewak}
\author[4]{Marcin Kowalski}
\author[2]{Marcin Gębski}
\author[3]{Marcin Motyka}
\author[1]{Anna Szerling}
\author[2,*]{Tomasz Czyszanowski}

\affil[1]{Łukasiewicz Research Network - Institute of Microelectronics and Photonics, al. Lotników 32/46, 02-668 Warsaw, Poland}
\affil[2]{ Photonics Group, Institute of Physics, Lodz University of Technology, ul. Wolczanska 219, 90-924 Łódź, Poland}
\affil[3]{Laboratory for Optical Spectroscopy of Nanostructures, Department of Experimental Physics, Faculty of Fundamental Problems of Technology, Wrocław University of Science and Technology, Wybrzeże Wyspiańskiego 27, 50-370 Wrocław, Poland}
\affil[4]{Institute of Optoelectronics, Military University of Technology, Gen. S. Kaliskiego, 00-908 Warsaw, Poland}
\affil[*]{Corresponding author: tomasz.czyszanowski@p.lodz.pl}

\begin{document}
\twocolumn[
\begin{@twocolumnfalse}
\maketitle
\begin{abstract}
  The design of transparent conductive electrodes (TCEs) for optoelectronic devices requires a trade-off between high conductivity or transmittivity, limiting their efficiency. This paper demonstrates a novel approach to fabricating TCEs that effectively alleviates this trade-off: a monolithic GaAs high contrast grating integrated with metal (metalMHCG). The metalMHCG enables higher electrical conductivity than other TCEs, while providing transmissive and antireflective properties. We focus on infrared spectrum TCEs, which are essential for sensing, thermal imaging, and automotive applications. However, due to elevated free carrier absorption they are much more demanding than TCEs for the visible spectrum. We demonstrate 75\% absolute transmittiance of unpolarized light, resulting in 108\% transmittance relative to plain GaAs substrate. We achieved even larger absolute transmittance of polarized light, reaching 92\% or 133\% relative transmittance. Despite record high transmittance, the sheet resistance of the metalMHCG is several times lower than any other TCE, ranging from 0.5 to \SI{1}{\OSq}.  
\newline
\newline
 \end{abstract}
 \end{@twocolumnfalse}
]

%\begin{multicols}{2}
\section{Introduction}

Optoelectronic technology relies on transparent conductive electrodes (TCEs) enabling electrical conductivity within the electrode plane, current injection to the semiconductor on which the TCE is deposited, and transmittance of light through the TCE. The high concentration of free carriers in TCEs is the main limiting factor affecting the performance of TCEs and imposes a fundamental trade-off between electrical conductivity and light absorption (accompanied by reflection of light propagating through the TCEs, in accordance with the Drude model). 

Visible light transparent conductive oxides (TCOs) are widely used as TCEs in liquid crystal displays \cite{Lagerwall_Scalia_2012}, light-emitting diodes \cite{Gather_Kohnen_Meerholz_2010}, and solar cells \cite{Yang_Liu_Xiong_Zhang_Liang_Yang_Xu_Ye_Chen_2014, Rider_Tucker_Worfolk_Krause_Lalany_Brett_Buriak_Harris_2011}. The most common TCO is indium tin oxide (ITO). However, despite its excellent performance, ITO has some significant drawbacks, including high cost, brittleness, recycling difficulties, and the scarcity of indium \cite{Kumar_Zhou_2010, Bel_Hadj_Tahar_Ban_Ohya_Takahashi_1998}. To overcome these issues, other TCOs, such as fluorine-doped tin oxide (FTO) and aluminum-doped zinc oxide (AZO), have emerged as attractive alternatives to ITO \cite{Koo_Oh_Riu_Ahn_2017}.  Not only do  they do not rely on indium, but they can achieve a high level of absolute transmittance of up to 90\%, which is comparable to ITO. Their sheet resistance levels are typically above 10 \SI{}{\OSq} for transmittance levels of around 80\%, which is also similar to ITO.  

Other alternatives include ultrathin metals \cite{Xu_Shen_Cui_Wen_Xue_Chen_Chen_Zhang_Li_Li_et_al._2017, Bi_Liu_Zhang_Yin_Wang_Feng_Sun_2019, Surya_Prakasarao_Hazarika_DSouza_Fernandes_Kovendhan_Arockia_Kumar_Paul_Joseph_2020, Chueh_Chien_Yip_Salinas_Li_Chen_Chen_Chen_Jen_2012}, conductive polymers \cite{Anand_Islam_Meitzner_Schubert_Hoppe_2021, Kim_Kee_Lee_Lee_Kahng_Jo_Kim_Lee_2013, Vosgueritchian_Lipomi_Bao_2011}, carbon nanotubes \cite{Jeon_Delacou_Okada_Morse_Han_Sato_Anisimov_Suenaga_Kauppinen_Maruyama_et_al._2018, Daneshvar_Tagliaferri_Chen_Zhang_Liu_Sue_2020}, graphene-based layers \cite{Liu_You_Liu_Yan_2015,Shin_Jang_Lee_Seo_Choi_2018,Kang_Choi_Park_Park_Cho_Cho_Walker_Choi_Shin_Seo_2021, Huang_Sheng_Tu_Zhang_Wang_Geng_Zou_Di_Yi_Sun_et_al._2015, Wang_Liu_Pan_Bai_Xie_Zhao_Xue_Wen_Chen_2022}, nanowires \cite{Lee_Connor_Cui_Peumans_2008, Guo_Lin_Chen_Wang_Xie_Zheng_Gao_Li_Kang_Cai_et_al._2013, Hsu_Kong_Wang_Wang_Welch_Wu_Cui_2014, Guo_Li_Radmilovic_Radmilovic_Turbiez_Spiecker_Forberich_Brabec_2015}, dielectric-metal-dielectric structures \cite{Cattin_El_Mahlali_Cherif_Touihri_El_Jouad_Mouchaal_Blanchard_Louarn_Essaidi_Addou_et_al._2020, Acosta_Mendez-Gamboa_Riech_Acosta_Zambrano_2019, Cattin_Louarn_Morsli_Bernede_2021, Mouchaal_Louarn_Khelil_Morsli_Stephant_Bou_Abachi_Cattin_Makha_Torchio_et_al._2015,Kim_Won_Woo_Jeong_Moon_2014,Ji_Liu_Zhang_Jay_Guo_2020}, and metal networks \cite{Rao_Hunger_Gupta_Kulkarni_Thelakkat_2014, Wu_Kong_Ruan_Hsu_Wang_Yu_Carney_Hu_Fan_Cui_2013}. Of these alternatives, metal nanowires and dielectric-metal-dielectric structures are the most promising, as they exhibit low sheet resistance below 10 \SI{}{\OSq} and above 90\% absolute transmittance. Graphene is also highly attractive, due to its theoretical potential for high conductivity \cite{Fang_Konar_Xing_Jena_2007} and high transmittance, which can reach 95\% for the visible light spectral range in the case of a single graphene monolayer. However, current methods for synthesizing graphene films have resulted in resistance values that are too high for electrodes used in electronic devices \cite{Ye_Yan_Xie_Kong_Liang_Chen_Xu_2017}. 

The implementation of TCEs directly on semiconductors  is highly desirable for electroluminescent diodes, photodiodes, lasers, and other technologies that predominantly work in the infrared spectral range and where high current density is injected into the devices. The utilization of TCEs in infrared semiconductor devices is impacted by two phenomena that negatively affect their transmittance properties. Firstly, the plasma frequency of free electrons coincides with the frequency of infrared light, which enhances the response of free electrons in the TCEs to electromagnetic excitation. This heightened response leads to increased absorption and reflection, and as a result the transmittivity of infrared TCEs is significantly lower \cite{Krause_Miclea_Steudel_Schweizer_Seifert_2014, Tikuišis_Dubroka_Uhlířová_Speck_Seyller_Losurdo_Orlita_Veis_2023,Massonnet_Carella_Jaudouin_Rannou_Laval_Celle_Simonato_2014} compared to TCEs used in the visible range. Recently, S. Wang et al. \cite{Wang_Liu_Pan_Bai_Xie_Zhao_Xue_Wen_Chen_2022} demonstrated record high infrared light transmittance and low sheet resistance using conductive composite films based on silver nanowires and graphene implemented on a $\textup{BaF}_2$ substrate, achieving transmittance over 70\% and sheet resistance of \SI{28}{\OSq}. However, there is a general lack of research on TCEs in the infrared region, despite their wide range of possible applications in devices for thermal imaging, free space communication, eye-safe LIDARs, and the detection of toxic gases—which in the wavelength range of 4 to \SI{10}{\micro\meter} have strong absorption lines. 

Secondly, the transmittivity of TCEs implemented on the interface between air and substrate of high refractive index is limited by Fresnel reflection, given by  the formula: 

\begin{equation}
    R=\left(\frac{n-1}{n+1}\right)^2
\end{equation}
where $n$ is the refractive index of the substrate and the refractive index of air is 1. In the case of glass and polymer substrates, Fresnel reflection is not larger than 5\%. The interface between air and most narrow-bandgap semiconductors defined by refractive indices close to 3 introduces reflection of nearly 30\%. Thus, if the TCEs discussed earlier are deposited on the surface of a semiconductor, their transmittance will be lower by $\sim$25\%. 

We define the relative transmittance of a TCE as the ratio between the absolute transmittance of the TCE on a transparent substrate and the absolute transmittance of the same substrate without the TCE. This measure allows us to compare the optical properties of TCEs employed for various purposes. Relative transmittance enables determination of the transmittance regardless of the impact of Fresnel reflection. The relative transmittance of the TCEs discussed so far in this article remains below 100\%. This is due mostly to the absorption and reflection induced by free carriers. However, it has been demonstrated that by leveraging low quality factor resonance in an $\textup{Al}_2\textup{O}_3$/Cu:Ag/ZnO planar configuration, it is possible to achieve a relative transmittance slightly above 100\% within the visible light range, while maintaining a sheet resistance of \SI{18}{\OSq} \cite{Ji_Liu_Zhang_Jay_Guo_2020}.   

\begin{figure*}[h!]
\centering
\includegraphics[width=0.7\textwidth]{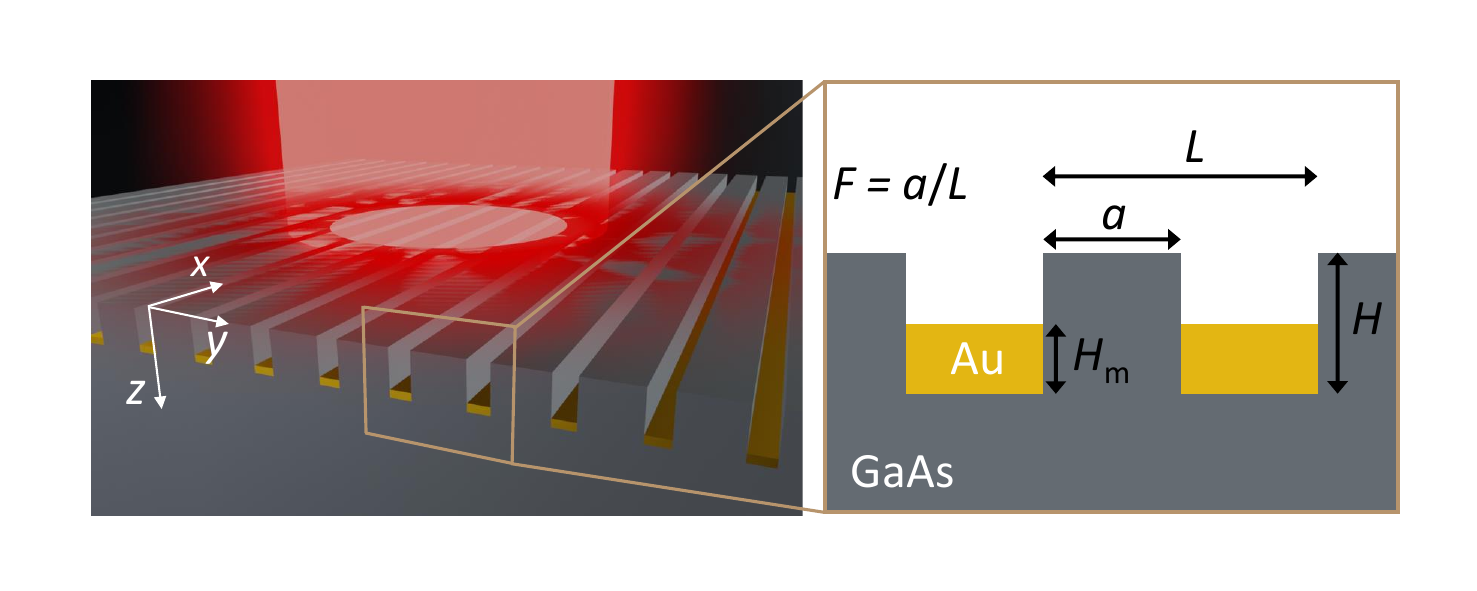}
\caption{\label{fig:1}Schematics of the metalMHCG configuration composed of one-dimensional grating on a GaAs wafer with gold stripes implemented in the grooves between the semiconductor stripes. The geometrical parameters of the grating and the coordinate system are indicated.}
\end{figure*}

To address the inherent trade-off between resistivity and transmittivity in TCEs, extend their spectral range of transparency, and mitigate the Fresnel reflection, in \cite{Czyszanowski_Sokol_Dems_Wasiak_2020} we proposed two configurations of semiconductor deep-subwavelength monolithic high-contrast gratings integrated with metal (metalMHCG). The two configurations, named TM and TE, are composed of a one-dimensional semiconductor grating integrated with metal stripes on bulk semiconductor material matching the grating. In the TM configuration metal stripes are implemented on the semiconductor stripes, while in the TE configuration metal stripes are deposited between semiconductor stripes (see Fig. \ref{fig:1}). The nomenclature of the configurations is derived from the predominant polarization component of light that is transmitted by them. As such, the TM configuration primarily transmits the TM component of the electromagnetic field, where the electric field of the electromagnetic wave oscillates perpendicular to the stripes in the grating plane (along the $y$-axis, see Fig. \ref{fig:1}). The TE configuration predominantly transmits the TE component, where the electric field oscillates along the grating stripes (along the $x$-axis, see Fig. \ref{fig:1}). In what follows, TE and TM transmittances refer to the transmittances of the TE and TM light components, respectively. Unpolarized light consists of equal TE and TM components. The metalMHCG may partially polarize the incident unpolarized light under transmission. However, to characterize the metalMHCG in terms of its ability to transfer the energy of unpolarized light regardless of the polarizing effect of the transmission, we define the transmittance of unpolarized light as the average of the TE and TM components that are transmitted, bearing in mind that the transmitted light may be partially polarized. 

In \cite{Czyszanowski_Sokol_Dems_Wasiak_2020}, we showed by numerical analysis that metalMHCGs in both configurations can achieve absolute transmittance of polarized light of more than 98\% (corresponding to a relative transmittance of more than 140\%) for electromagnetic radiation ranging from the visible to far infrared spectrum, while their sheet resistance can be less than \SI{1}{\OSq}. The high transmittance of the metalMHCG in the visible and infrared ranges is made possible by the low quality factor resonance occurring in the air slits for TM polarization and in the semiconductor stripes of the grating for TE polarization, resulting in reduced light interaction with the metal stripes \cite{Sokol_Czyszanowski_2020}. Thus, the minimized electron excitation by the electromagnetic wave in metal sustains high transmittance, even when the frequency of light approaches the electron plasma frequency. 

\begin{figure*}[h!]
\centering
\includegraphics[width=1.0\textwidth]{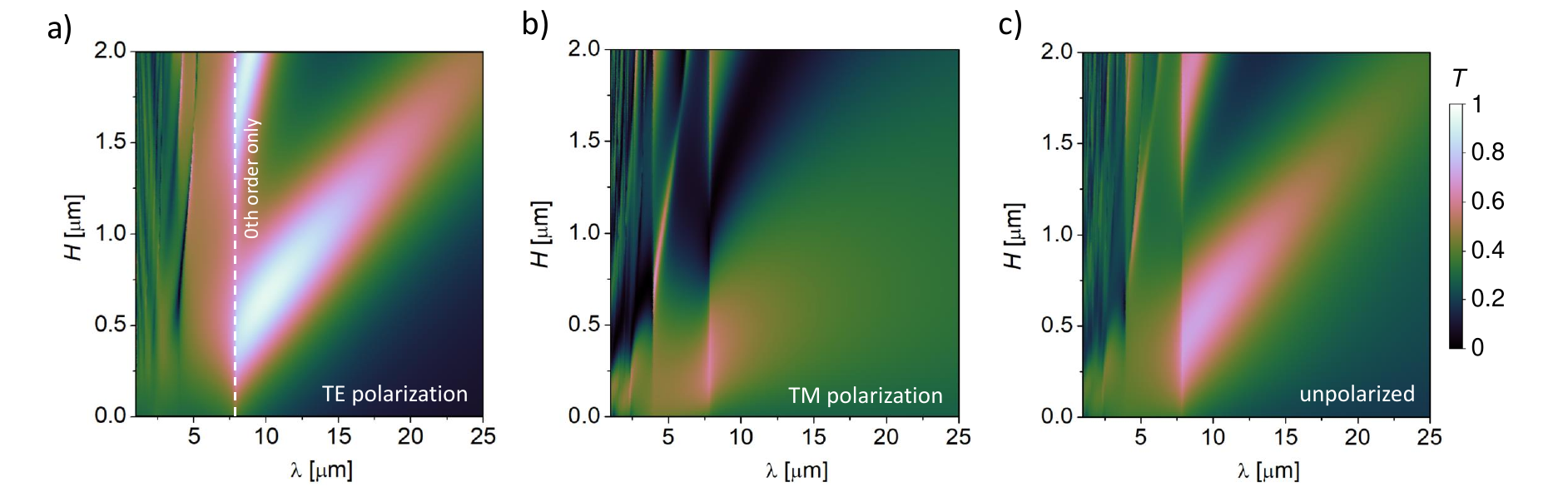}
\caption{\label{fig:2}Transmittance ($T$) maps of metalMHCG under normal incidence of TE polarized a), TM polarized b) and unpolarized light c) in the domain of the wavelength ($\lambda$), and height of the semiconductor stripes ($H$) for $L = \SI{2.4}{\micro \meter}$, $F = 0.7$, $H_\textup{m} = \SI{0.2}{\micro \meter}$. The white dashed line in a) indicates the limit of the zeroth diffraction order emitted towards substrate. The refractive indices of GaAs and Au are based on values reported in \cite{Skauli_Kuo_Vodopyanov_Pinguet_Levi_Eyres_Harris_Fejer_Gerard_Becouarn_et_al._2003, Ordal_Bell_Alexander_Long_Querry_1987}}
\end{figure*}

In a previous study \cite{Tobing_Wasiak_Zhang_Fan_Czyszanowski_2021}, we experimentally demonstrated 90\% transmittance of polarized light through a metalMHCG in the TM configuration with a corresponding sheet resistance of \SI{2}{\OSq}. This configuration showed strong discrimination of the orthogonal polarization. According to the numerical analysis, it is evident that the TE configuration, as shown in Fig. \ref{fig:1}, exhibits lower polarization discrimination compared to the TM configuration. In another study \cite{Sokol_Czyszanowski_2020}, we showed by numerical analysis that with an optimised design of the TE configuration absolute unpolarized transmittance as large as 95\% can be achieved, corresponding to 137\% relative transmittance. However, the optimised configuration for unpolarized light transmittance requires that the height of the semiconductor stripes is twice the period of the grating and the width of the grooves between the stripes is about 20\% of the period, which poses a major technological challenge.

Here, we present the experimental realization of a metalMHCG in the TE configuration that is designed to maximize transmittance of TE polarized light in the vicinity of the \SI{9}{\micro\metre} wavelength. The metalMHCG achieved 92\% absolute transmittance and 133\% relative transmittance of TE polarized light compared to a bare GaAs substrate. The transmittance bandwidth of polarized light above 100\% of the relative transmittance is wider than \SI{3}{\micro\metre}. Since in this configuration the metalMHCG also enables high transmittance of the orthogonal light component (TM), the absolute transmittance of non-polarized light reaches 75\%. This corresponds to 108\% relative transmittance and \SI{2}{\micro\metre} bandwidth above 100\% of the relative transmittance, even though the metalMHCG is not designed for unpolarized light transmittance as in \cite{Sokol_Czyszanowski_2020}. Additionally, our metalMHCGs are characterised by sheet resistance of 0.5-\SI{1}{\OSq}, which is several times lower than any TCE reported previously. The metalMHCG can be monolithically integrated with a wide range of materials used in optoelectronics, particularly with high refractive index semiconductors, effectively eliminating surface reflection. Additionally, the weak interaction between light and metal in the metalMHCG minimizes metal reflection and absorption. These unique characteristics allow for substantially higher transmittance and metal content compared to other TCE realizations. As a result, our electrode exhibits low sheet resistance while maintaining high transmittance, even in the challenging infrared electromagnetic spectrum.

\section{Results}
The metalMHCG in the TE configuration is composed of a monolithic, double-side polished \SI{350}{\micro \metre} thick undoped GaAs wafer with parallel stripes etched onto one surface. Gold stripes are deposited in the grooves between the semiconductor stripes (see Fig. 1). The back side of the wafer is not covered with an anti-reflective coating, as most dielectrics have high absorption in the spectral range of interest. The parameters defining the metalMHCG geometry are as follows: $L$ – period of the grating; $H$, $H_\textup{m}$ – heights of the semiconductor and metal stripe, respectively; $a$, $a_\textup{m}$  – widths of the semiconductor and metal stripe respectively, where $a + a_\textup{m} = L$; $F$ – the duty cycle, which is the ratio of the width of the semiconductor stripe ($a$) to the period ($L$); $n_\textup{s}$, $n_\textup{m}$ – the refractive indices of the semiconductor and metal, respectively. In the numerical analysis that complements the experiment, we assume a semi-infinitely thick wafer and a semi-infinite air superstrate above the grating. We consider a single period of the grating with periodic boundary conditions, which elongates the metalMHCG to infinity in lateral directions. In this section, we will conduct an analysis of absolute and relative transmittances through numerical simulations and experimental measurements. The term “transmittance” will refer to absolute transmittance, whereas relative transmittance will be explicitly indicated. 

\textbf{Numerical design.} In this section, we elucidate the properties of the optimal metalMHCG structure, before delving into the process of its development. Figure \ref{fig:2} illustrates calculated transmittance maps for close to optimal metalMHCG in the TE configuration with a period ($L$) of \SI{2.4}{\micro \meter}, duty cycle ($F$) of 0.7, and thickness of gold ($H_\textup{m}$) of \SI{200}{\nano\metre}, in the domain of the wavelength ($\lambda$) and the height of the semiconductor stripes ($H$) for TE polarization (Fig. \ref{fig:2}a), TM polarization (Fig. \ref{fig:2}b), and unpolarized light (Fig. \ref{fig:2}c). High transmittance exists in the range $\lambda > Ln_\textup{s}$ . Its short wavelength border is indicated by the white dashed line in Fig. \ref{fig:2}a. In this region, called the deep-subwavelength region, only the zeroth diffraction order can be transmitted towards the air and substrate. Regions of high TE polarization transmittance (Fig. \ref{fig:2}a) repeat as $H$ increases \cite{Czyszanowski_Sokol_Dems_Wasiak_2020}, although this is not shown in the figure. We focus on the region with the smallest $H$ enabling high transmittance, due to the simplicity of implementation of low $H$ gratings. The region of transmittance of TE polarized light higher than 90\% ranges above \SI{5}{\micro\metre} with respect to wavelength and above \SI{0.5}{\micro\metre} with respect to $H$. The transmittance of orthogonal TM polarization of light through the same grating is also substantial, and reaches above 60\% in local maxima. Transmittance of the same polarization above 50\% is present in a broad range of $\lambda$ and $H$ (see Fig. \ref{fig:2}b). The transmittance of unpolarized light (Fig. \ref{fig:2}c) follows the pattern of transmittance in the TE polarization map (Fig. \ref{fig:2}a), indicating very high transmittance above 70\% in a large area in the ($\lambda, H$) domain. It is noteworthy that although the configuration is not specifically optimized for maximal transmittance of unpolarized light, it still exhibits a remarkable level of unpolarized light transmittance that cannot be achieved by the metalMHCG in the TM configuration analyzed in \cite{Tobing_Wasiak_Zhang_Fan_Czyszanowski_2021}. 

The electrical properties of the metalMHCG can be characterised by sheet resistance, estimated using the formula:
\begin{equation} \label{eq2:2}
    R_\textup{s}=\frac{\rho L}{H_\textup{m}a_\textup{m}}=\frac{\rho}{H_\textup{m}(1-F))}
\end{equation}
where $\rho$ is the electrical resistivity of the gold stripes. The second formula is valid in the case of rectangular cross-section of the semiconductor stripes. To accurately model the resistivity of gold stripes with nanoscale cross-sectional dimensions, it is necessary to consider the resistivity-size effect, which increases the electrical resistivity of the metalMHCG stripes above the level of the resistivity of bulk gold ($2.4\cdot10^{-8}$ \SI{}{\ohm \meter}) \cite{Gilani_Rabchuk_2018}. In numerical calculations, we assume the resistivity of the golden stripes of the metalMHCG to be equal to the resistivity of bulk gold. Doing so, we determine the possible bottom limit of the metalMHCG sheet resistivity. It is important to note that in a realistic case metal stripes are expected to have lower conductivity compared to their bulk counterpart, due to possible impurities incorporated during fabrication, reduced crystal quality, contaminants or surfactants on the surface, and electron scattering. Applying formula \ref{eq2:2}, the sheet resistance of the configuration analysed in Fig. \ref{fig:2} is around {\SI{0.4}{\OSq}}. 

We shall now describe the methodology employed to find the optimal structure. To achieve maximal transmittance of TE polarized light, meticulous integration of the metal stripes with a semiconductor grating is necessary. The metal stripes should possess sufficient thickness to induce a waveguide effect, enabling the effective funnelling of  radiation through the semiconductor stripes \cite{Czyszanowski_Sokol_Dems_Wasiak_2020,Sokol_Czyszanowski_2020}. On the other hand, excessive thickness of the metal stripes reduces the transmittance, due to absorption. Considering the interplay of these opposing effects, a global maximum in the transmittance of TE-polarized light can be attained by adjusting the quantity of metal in the metalMHCG structure.  

\begin{figure*}[h!]
\centering
\includegraphics[width=1.0\textwidth]{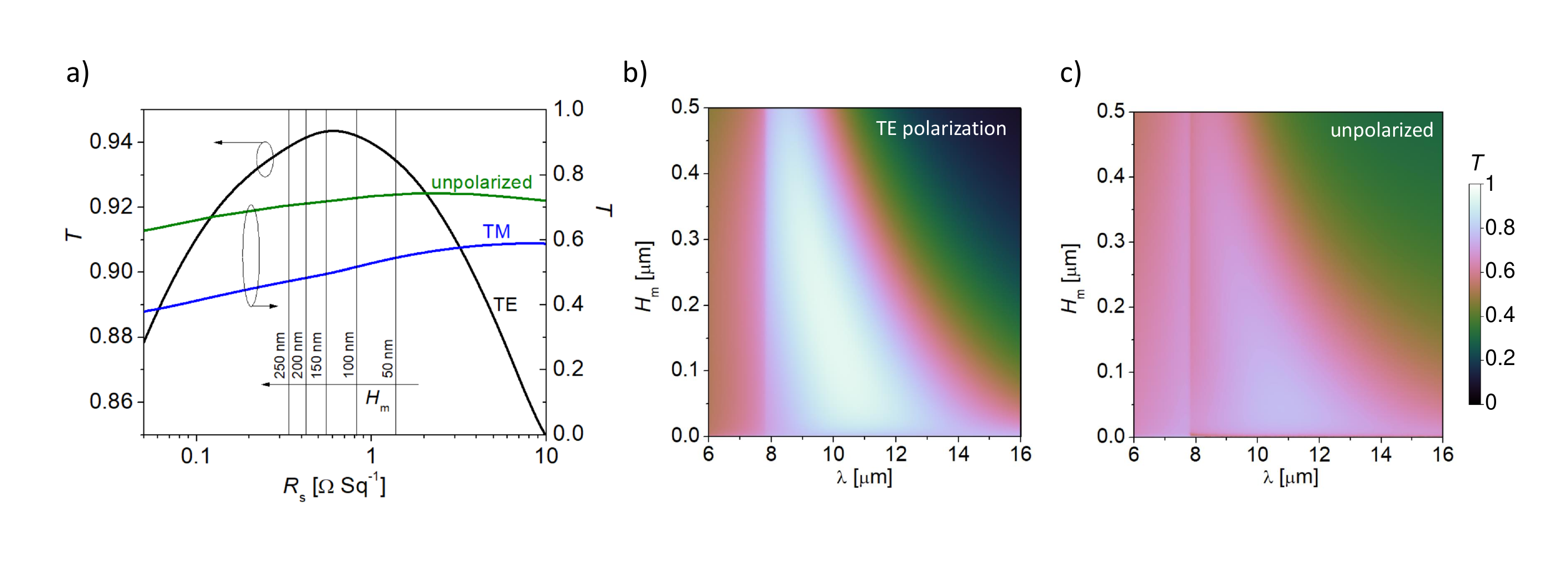}
\caption{\label{fig:3}a) MetalMHCG maximal transmittance for TE polarization (black curve, left vertical axis) and corresponding transmittance of TM polarized (blue curve, right vertical axis) and unpolarized light (green curve, right vertical axis) as a function of the sheet resistance ($R_\textup{s}$) of golden stripes. The black vertical lines indicate the thicknesses of the gold layers in the metalMHCG realised in the experimental part of this work. Maps of TE polarized b) and unpolarized light c) transmittance in the domain of the wavelength ($\lambda$) and thickness of the gold stripes ($H_\textup{m}$) for metalMHCG with parameters $L =\SI{2.394}{\micro \meter}$, $F= 0.71$, $H = \SI{0.655}{\micro \meter}$.}
\end{figure*}

We employ the multidimensional Nelder-Mead simplex algorithm \cite{Wright} to search for the maximal transmittance of TE polarized light for a given value of metalMHCG sheet resistance. The variables considered in the optimization process are $L, F, H, H_\textup{m}$. However, $L, F, H_\textup{m}$ are interrelated through Formula \ref{eq2:2}. In Fig. \ref{fig:3}a, each point on the black curve represents a different configuration of the metalMHCG, obtained by maximizing the transmittance of TE polarized light. The TM polarized and unpolarized light transmittances corresponding to the metalMHCG configurations are indicated by blue and green curves in the same figure. The black curve indicates that the global maximum of TE polarized light transmittance reaches 94.3\%, which corresponds to sheet resistance of \SI{0.58}{\OSq}. The transmittance of unpolarized light reaches a maximum value of approximately 74.2\% for $R_\textup{s}$ of \SI{2}{\OSq}. As the value of $R_\textup{s}$ decreases, the transmittance of unpolarized light gradually decreases, reaching 70\% for $R_\textup{s} =\SI{0.28}{\OSq}$. In the experimental analysis, our goal is to achieve the lowest possible sheet resistance while maintaining transmittance values close to the maximum. Therefore, we choose as the geometrical parameters of our target metalMHCG configuration $L =\SI{2.394}{\micro \meter}$, $F = 0.71$, $H =\SI{0.655}{\micro \meter}$, and $H_\textup{m} =\SI{0.2}{\micro \meter}$, which theoretically allow TE transmittance of 94.1\%, unpolarised light transmittance of 71.1\% and $R_\textup{s}=\SI{0.42}{\OSq}$. Although this configuration exhibits a sheet resistance that is 40\% lower than the configuration providing maximum TE transmittance, the transmittance is only 0.2\% lower in comparison to the configuration with maximum TE transmittance. 

Figure \ref{fig:3}b maps the TE transmittance in the $\lambda$ and $H_\textup{m}$ domain for the target configuration. Increasing the thickness of the gold stripes while keeping $L$, $F$, and $H$ constant results in a narrowing of the high transmittance spectrum width and a shifting of the spectrum maximum towards shorter wavelengths with global maximum for $H_\textup{m}=\SI{0.2}{\micro \meter}$ The unpolarized transmittance shown in Fig. \ref{fig:3}c follows a similar pattern to the TE transmittance map however maximal unpolarized transmittance reduces slowly with the increase in $H_\textup{m}$. These trends will be observed in the forthcoming section on experimental implementation.

\textbf{Experimental demonstration.} Five configurations of the metalMHCG were fabricated by a combination of plasma enhanced chemical vapour deposition (PECVD), electron beam lithography (EBL), plasma etching by inductively coupled plasma-reactive ion etching (ICP-RIE), and e-beam physical vapour deposition (EBPVD). The configurations possess the same nominal grating parameters of $L=\SI{2.40}{\micro \meter}$, $F=0.71$, $H=\SI{0.65}{\micro \meter}$ and five different thickness of the gold stripes ranging from 0.05 to \SI{0.25}{\micro \meter}, with increments of \SI{0.05}{\micro \meter}. The process of metalMHCG fabrication is described in Supplementary material, Section \ref{sec:s1}. 

\begin{figure*}[h!]
\centering
\includegraphics[width=1.0\textwidth]{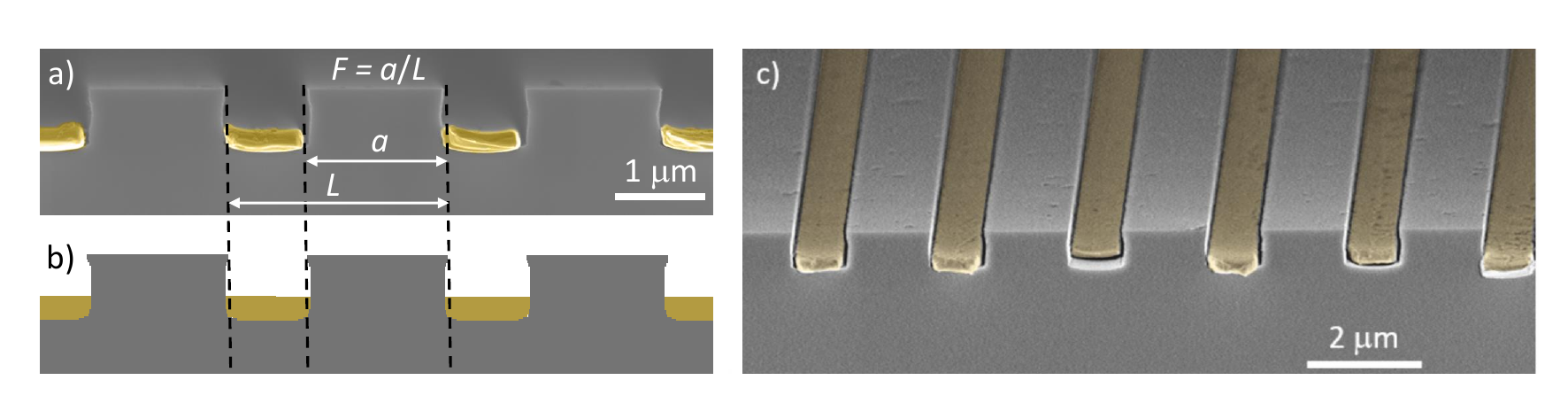}
\caption{\label{fig:4}a) Cross-section scanning electron microscope (SEM) image of metalMHCG; b) image of the metalMHCG implemented in the numerical algorithm; c) tilted SEM image of cleaved edge of metalMHCG. The thickness of the gold stripes is \SI{200}{\nano \meter} in all images. }
\end{figure*}

To prevent gold deposition on the semiconductor side walls, which could potentially reduce transmittance, the parameters of the ICP-RIE etching process were optimized to achieve a concave side wall profile, as illustrated in Fig. \ref{fig:4}a.  Such a cross-section of the semiconductor stripes requires new optimal grating parameters that are different from the parameters of a rectangular cross section. Based on scanned SEM images, the metalMHCG cross-section was incorporated into a numerical optimisation algorithm (Fig. \ref{fig:4}b) providing new dimensional parameters ($L= \SI{2.51}{\micro \meter}$, $F=0.68$, $H=\SI{0.71}{\micro \meter}$). Figures \ref{fig:4}a and \ref{fig:4}b illustrate the graphical definition of $F$ for the concave cross-section of the grating stripes. The TE and unpolarized transmittances for new grating parameters are nearly indistinguishable from the transmittance of metalMHCG with rectangular cross-sections for the same $L$, $F$, $H$ and $H_\textup{m}$.  

\begin{figure*}[h!]
\centering
\includegraphics[width=0.9\textwidth]{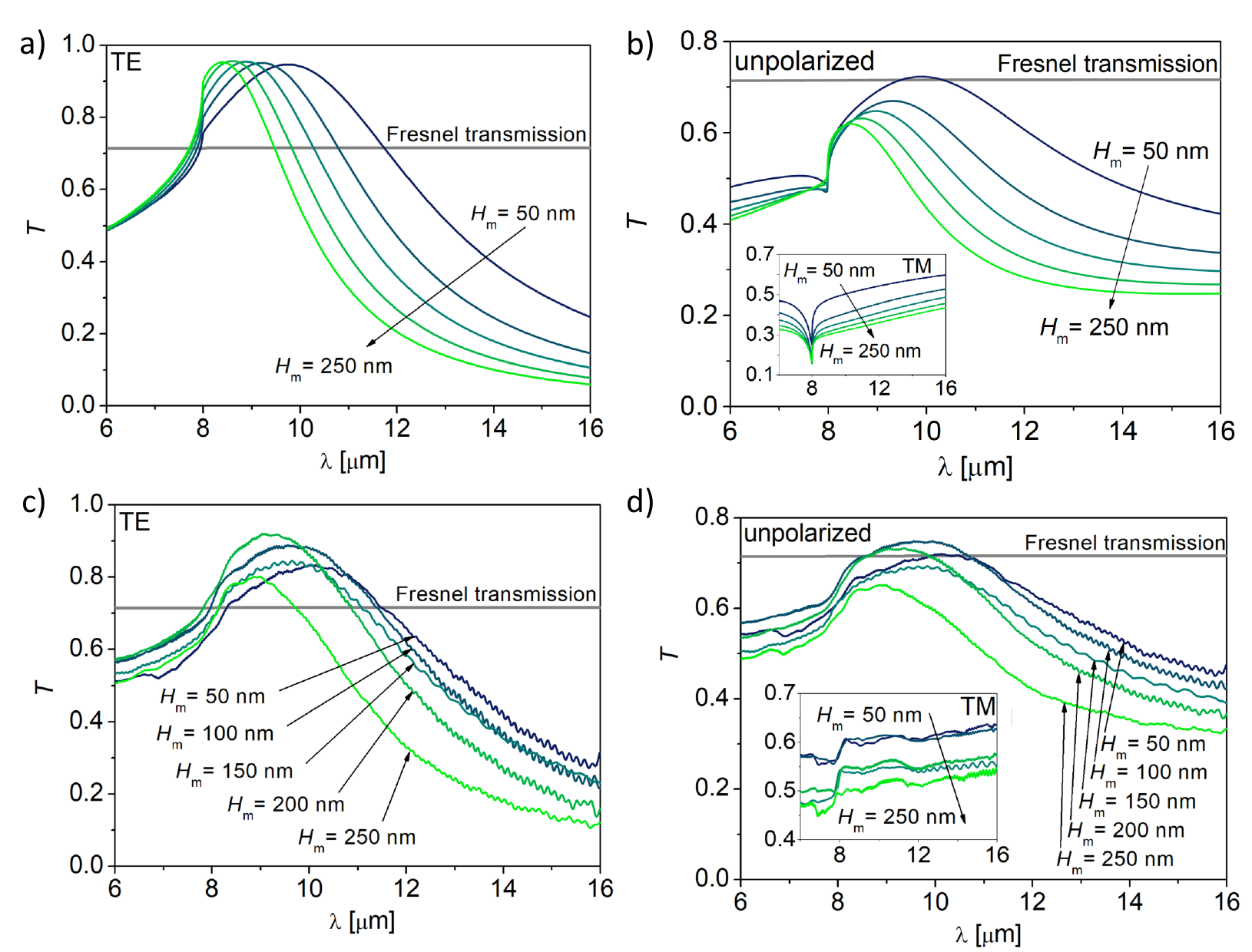}
\caption{\label{fig:5}Calculated a, b) and measured c, d) transmittance spectra of metalMHCG in the case of TE-polarized a, c) and unpolarized b, d) light incidence. The spectra are calculated for real-world cross-sections of semiconductor and metal stripes with the geometrical parameters of the metalMHCG determined by SEM images ($L =\SI{2.53}{\micro \meter}$, $F = 0.67$, and $H =\SI{0.62}{\micro \meter}$) and $H_\textup{m}$ from \SI{50}{\nano \meter} to \SI{250}{\nano \meter} indicated by different colours. The same colours correspond to measured spectra of samples with the same thickness of gold stripes. The grey line represents Fresnel transmittance through the planar interface between the GaAs and air, based on GaAs refractive index dispersion \cite{Skauli_Kuo_Vodopyanov_Pinguet_Levi_Eyres_Harris_Fejer_Gerard_Becouarn_et_al._2003}. The insets in b) and d) illustrate TM transmittance of the metalMHCG.}
\end{figure*}

The actual cross-section shape and dimensions of the processed metalMHCG were determined based on SEM images. All five samples with nominally different metal stripe thickness had the same geometrical parameters of the semiconductor grating: $L =\SI{2.53}{\micro \meter}$, $F = 0.67$, and $H =\SI{0.62}{\micro \meter}$, determined with a precision of \SI{10}{\nano \meter} (comparison of calculated transmittance spectra of optimal configuration and realised are depicted in Suplemetary meterial  Fig. \ref{fig:OptvsRe}). The calculated spectra of TE transmittance (see Fig. \ref{fig:5}a) exhibit maxima above 90\% for the parameters and cross section of the processed metalMHCG. When the thickness of the gold stripes is increased from \SI{50}{\nano \meter} to \SI{250}{\nano \meter}, the wavelength of the calculated TE transmittance maxima shifts towards shorter wavelengths by \SI{1.34}{\micro \meter}. The metalMHCG with \SI{200}{\nano \meter} thick gold stripes has a maximal calculated TE transmittance of 95.6\% at a wavelength of \SI{8.6}{\micro \meter}. The calculated unpolarized transmittance (Fig. \ref{fig:5}b) follows the behaviour of TE transmittance, as the spectrum of TM transmittance is a slowly varying function of wavelength in the spectral range under consideration, as seen in the inset of Fig. \ref{fig:5}b. The maximum of calculated unpolarized transmittance is achieved for metalMHCG with $H_\textup{m}=\SI{50}{\nano \meter}$ and reaches above 70\% at a wavelength of \SI{10}{\micro \meter}. As the thickness of the metal increases, the transmittance of unpolarized light decreases, mainly due to the decrease in TM transmittance (see inset in Fig. \ref{fig:5}b). 

Figure \ref{fig:5}c presents the measured TE transmittance spectra for the five metalMHCG samples. The description of the experimental setup for transmittance measurements and the experimental method are detailed in Supplementary material, Section \ref{sec:s2}. The spectra show a maximum that shifts towards shorter wavelengths by \SI{1.21}{\micro \meter} as the gold stripe thickness increases from \SI{50}{\nano \meter} to \SI{250}{\nano \meter}. The maximal TE transmittance is achieved for a metalMHCG with \SI{200}{\nano \meter} thick gold stripes, reaching 92\% at a wavelength of \SI{9.08}{\micro \meter} which corresponds to relative TE transmittance of 133\%. The bandwidth of relative transmittance exceeding 100\% is \SI{3.03}{\micro \meter}. In the deep-subwavelength regime corresponding to wavelengths longer than \SI{8}{\micro \meter}, the TM transmittance is from 50\% to slightly above 60\% and varies slightly (see the inset in Fig. \ref{fig:5}d). Figure \ref{fig:5}d shows the unpolarized transmittance spectra for the five considered samples. The configuration with a gold stripe thickness of \SI{100}{\nano \meter} shows a maximum level of 75\%, which corresponds to relative transmittance of 108\% with an above 100\% relative transmittance bandwidth of \SI{2.03}{\micro \meter}. The discussed properties of the metalMHCG experimental spectra are in good agreement with the simulations. The inconsistencies, mainly concerning the level of transmittance, are typically related to the fabrication precision influencing grating stripes roughness. Therefore, the experimental transmittance can be enhanced by improving the process of metalMHCG fabrication.  
\begin{figure*}[h!]
\centering
\includegraphics[width=0.8\textwidth]{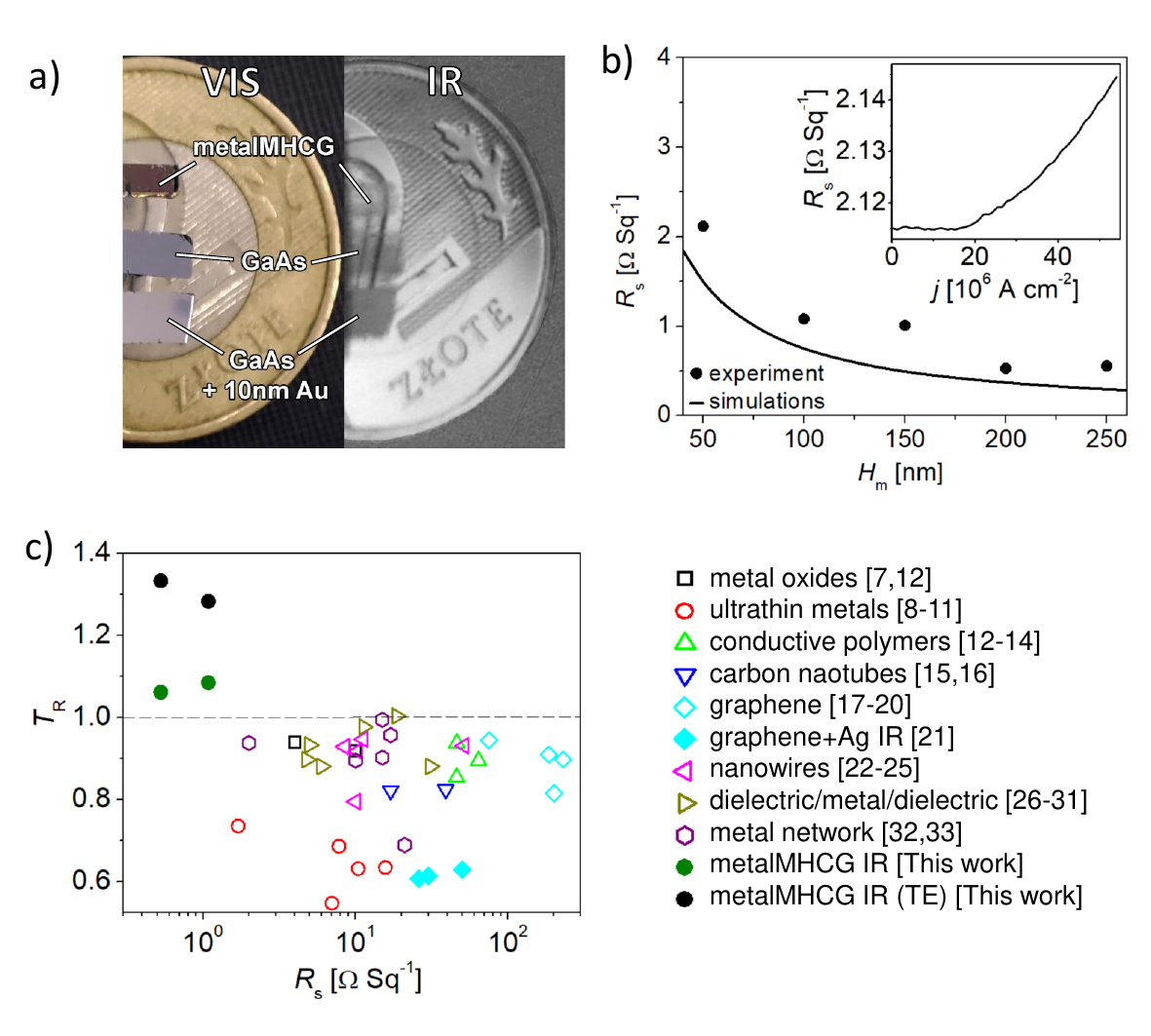}
\caption{\label{fig:6} a) Pictures of three samples from the same GaAs wafer: metalMHCG on the GaAs substrate (metalMHCG), bare GaAs substrate (GaAs), and GaAs substrate with a 10-nm thick gold layer on the surface (GaAs + \SI{10}{\nano \meter} Au). The pictures are shown in visible light (VIS) and in long-wavelength infrared (IR), with a coin in the background heated to 48\textdegree C, corresponding to the spectral maximum of black body radiation at a wavelength of \SI{9}{\micro \meter}; b) measured sheet resistance of the metalMHCG (indicated by dots) with varying thicknesses of gold stripes from \SI{50}{\nano \meter} to \SI{200}{\nano \meter} with \SI{50}{\nano \meter} steps. The simulated sheet resistance for a corresponding range of gold stripe thicknesses is represented by the black line. Inset illustrates measured dependence of sheet resistance versus current density; c) TE polarized (black dots) and unpolarized (green dots) optical relative transmittance ($T_\textup{R}$) versus sheet resistance for the metalMHCG with the best performance (from this work). The performances of TCEs based on other approaches are indicated by all other symbols and colors. References to works describing TCEs and their characteristics are indicated in the legend. The full symbols represent relative transmittance at a wavelength of $\sim\SI{9}{\micro \meter}$, the open symbols represent maximal relative transmittance in the visible range.}
\end{figure*}
The experimental unpolarized transmittance is slightly larger than the theoretical one. We hypothesise that this may be caused by the nonperfect flatness of the gold stripes walls reducing possible plasmonic effects and hence reducing absorption of TM polarization and contributing to higher transmittance. Overall, our experimental results show very good agreement with the calculations and confirm the feasibility of high-power transmittance using a metalMHCG. Photographs of the fabricated metalMHCG on GaAs substrate, along with plain GaAs substrate and a 10-nm thick gold layer on GaAs substrate, are presented in Fig. \ref{fig:6}a. The high transparency of the metalMHCG is evident in infrared light (IR), surpassing the transmission through the plain GaAs substrate. On the other hand, the 10-nm thick gold layer on the GaAs substrate enables only 2\% transmittance, which is nontransparent to infrared light—as is clearly evident in the photography.

Determination of metalMHCG sheet resistance ($R_\textup{s}$) required the fabrication of separate samples. The electrical resistance of the metalMHCG samples used  for optical characterisation was significantly below the error of the measurement setup. To minimize the characterisation error, we designed the resistance of the metalMHCGs to be $\sim\SI{100}{\ohm}$. The resistances of the samples were at least three orders of magnitude higher than the measurement error. The samples made for electrical characterisation consisted of 12, 6, 4, 4, and 2 stripes with different gold thicknesses from \SI{50}{\nano \meter} to \SI{250}{\nano \meter} with a \SI{50}{\nano \meter} step and the stripes were \SI{2}{\milli \meter} in length. An example configuration is presented in Supplementary material, Fig. \ref{fig:S3}. The gold stripes were placed on an undoped GaAs substrate. The electrical connection was facilitated by electrical pads, positioned at both ends of the stripes. Based on the current-voltage characteristics and geometric dimensions of the gold stripes, the sheet resistance was determined for all samples. The sheet resistance is shown in Fig. \ref{fig:6}b, along with the theoretical values calculated assuming the bulk gold resistance. The sheet resistances of the metalMHCGs that exhibited the maximal unpolarized and TE transmittances, with 100 nm and 200 nm gold stripe thicknesses, were \SI{1}{\OSq} and \SI{0.5}{\OSq} respectively. The measured resistance of the gold stripes was approximately $3.5\cdot10^{-8}$ \SI{}{\ohm \meter}, which is 45\% larger than the bulk gold conductivity. The dependence of sheet resistance on the current flowing through metalMHCG was also determined. The generated Joule heat increases the temperature of the stripes, which increases the sheet resistance. The dependence of sheet resistance on current density for the metalMHCG with 50-nm thick gold stripes is shown in the inset of Fig. \ref{fig:6}b. A noticeable increase in sheet resistance occurs at current densities above $2\cdot10^7$ \SI{}{\ampere \per \centi \meter \squared}. An increase in sheet resistance can be expected at higher current densities for samples with thicker gold stripes. 

A more general comparison of the optical and electrical performance of the metalMHCG and numerous configurations of TCEs reported so far is presented in Fig. \ref{fig:6}c. If the relative transmittance of TCEs from other studies were not directly indicated, they were derived by taking into account their experimentally measured absolute transmittance and the refractive index of the substrates. Among the various TCEs, $\textup{Al}_2\textup{O}_3$-Cu:Ag-ZnO \cite{Ji_Liu_Zhang_Jay_Guo_2020} and Cu nanotrough networks \cite{Wu_Kong_Ruan_Hsu_Wang_Yu_Carney_Hu_Fan_Cui_2013} have exhibited the largest transmittivity to date, with their relative transmittance either slightly surpassing or approaching 100\%. In terms of electrical properties, the lowest sheet resistance has been achieved by a Cu nanotrough network\cite{Wu_Kong_Ruan_Hsu_Wang_Yu_Carney_Hu_Fan_Cui_2013}, FTO \cite{Anand_Islam_Meitzner_Schubert_Hoppe_2021}, dielectric-metal-dielectric electrodes composed of AZO-Ag-AZO \cite{Acosta_Mendez-Gamboa_Riech_Acosta_Zambrano_2019}, and $\textup{MoO}_3$-Ag-$\textup{MoO}_3$ \cite{Cattin_Louarn_Morsli_Bernede_2021}. These TCEs exhibit sheet resistances ranging from \SI{2}{\OSq} to \SI{6}{\OSq}. Conductive composite films based on silver nanowires and graphene measured in infrared range demonstrate relative transmittance over 60\% (absolute transmittance of 70\%) and sheet resistance of \SI{28}{\OSq}\cite{Wang_Liu_Pan_Bai_Xie_Zhao_Xue_Wen_Chen_2022}. The numerically calculated infrared transmittivities of TCEs implemented on high refractive index substrate are presented in Supplementary material, Fig. \ref{fig:S4}. The calculations reveal that TCEs based on ITO, AZO, FTO, metal plates, and grids reach the level of Fresnel transmittance when their sheet resistance exceeds \SI{100}{\OSq}. Notably, graphene layers demonstrate exceptional performance in this context, reaching Fresnel transmittance for sheet resistance significantly below \SI{100}{\OSq}.  \\
In this context, metalMHCGs exhibit substantial superiority over other TCEs in terms of both optical transmittance and electrical properties, as supported by theoretical and empirical evidence. The sheet resistance of the metalMHCGs is 4 times smaller compared to the best performing TCEs. Additionally, their polarized and unpolarized transmittances surpass the threshold of Fresnel transmittance, a feat that is difficult to achieve for other TCEs.  

\section{Discussion}
In this research, we performed an experimental demonstration of a monolithic high contrast grating integrated with metal stripes (metalMHCG). The high transmittance of transverse electric (TE) polarization through the metalMHCG  was enabled by the funnelling the light through semiconductor stripes that reduced reflection and absorption of the light by the metal stripes. This mechanism enabled transmittance of 92\% in our experiment, which corresponds to 130\% of relative transmittance. To our best knowledge, this is the record for transmittance demonstrated through a transparent conductive electrode (TCE). The structure shows \SI{3}{\micro \meter} bandwidth above 100\% relative transmittance for polarized light. The metalMHCG in the TE configuration that was considered in this work reveals weak polarization selectivity and enables significant transmittance of transverse magnetic (TM) polarization, which is enabled by the low-quality factor resonance located in the air slits between the semiconductor stripes. Transmittance of TM polarization is 50–60\% in a wide spectral range around maximal TE transmittance, leading to 75\% transmittance of unpolarized light corresponding to 108\% of relative transmittance. This is also the record value ever reported. The corresponding sheet resistance of the metalMHCG is \SI{1}{\OSq}. Another configuration achieved absolute and relative transmittances of 73\% and 106\%, respectively, and a record low sheet resistance of \SI{0.5}{\OSq}, surpassing the best experimental results and theoretical predictions for other TCEs over fourfold. All experimental results presented in this paper exhibit excellent agreement with theory, which predicted only a slightly different transmittance and 30\% -- 40\% lower sheet resistance compared to our experimental results. Nevertheless, it exhibits a remarkably low sheet resistivity, attributed to the unique capacity of the metalMHCG to accommodate a significantly larger volume of metal compared to any other TCE. The increased metal content does not compromise the light transmittance through the metalMHCG, due to the light funnelling mechanism which minimizes the interaction of the electromagnetic field with the metal and eliminates Fresnel reflection.

The underlying motivation for this research arose from the need for efficient methods for the development of transparent electrodes with the capability to inject high-density current into semiconductors across large electrode areas. This study presents a novel approach towards the realization of optically transparent and highly electrically conductive structures, which can be seamlessly integrated monolithically with various materials employed in optoelectronics, particularly those composed of high refractive index semiconductors. The metalMHCG configuration exhibits the ability to facilitate light transmittance spanning a broad spectral range, from ultraviolet to infrared \cite{Czyszanowski_Sokol_Dems_Wasiak_2020}. Achieving high transmittance of infrared light is particularly challenging due to much higher free carrier absorption in comparison to the visible range. The exceptional properties of the metalMHCG, coupled with the promise of high-density current injection into semiconductors pave the way for significant advancements in the efficiency of electroluminescent diodes, photodiodes, and semiconductor lasers. These properties also open new possibilities for innovative applications beyond the traditional uses of these devices. 

\subsection*{Acknowledgments}
MM acknowledges the support from the Polish National Science Center, grant OPUS 2019/33/B/ST7/02591. AS and ME acknowledge the support by the statutory funds of the Łukasiewicz Research Network – Institute of Microelectronics and Photonics. This work has been completed while WG and KB were the Doctoral Candidate in the Interdisciplinary Doctoral School at the Lodz University of Technology, Poland.

\bibliographystyle{unsrt}
\bibliography{Main_bibliography}
\appendix
\newpage

\onecolumn

\setcounter{equation}{0}
\setcounter{figure}{0}
\setcounter{section}{0}
\setcounter{table}{0}
\renewcommand{\thefigure}{S\arabic{figure}}
\renewcommand{\thesection}{S\arabic{section}}
\renewcommand{\theHtable}{Supplement.\thetable}
\renewcommand{\theHfigure}{Supplement.\thefigure}

\title{\textbf{Supplementary materials} \\ 
Monolithic High Contrast Grating Integrated with Metal: A Transparent Conductive Electrode for Infrared Optoelectronics with Exceptionally High Conductivity and Transmitivity}

\maketitle

This PDF file includes: 
\begin{itemize}
\item Supporting figures \ref{fig:S1}-\ref{fig:S4}
\item Supporting information \ref{sec:s1}, \ref{sec:s2}
\end{itemize}
\newpage

\section{Fabrication of metalMHCG}\label{sec:s1}
\begin{figure*}[h!]
\centering
\includegraphics[width=1.0\textwidth]{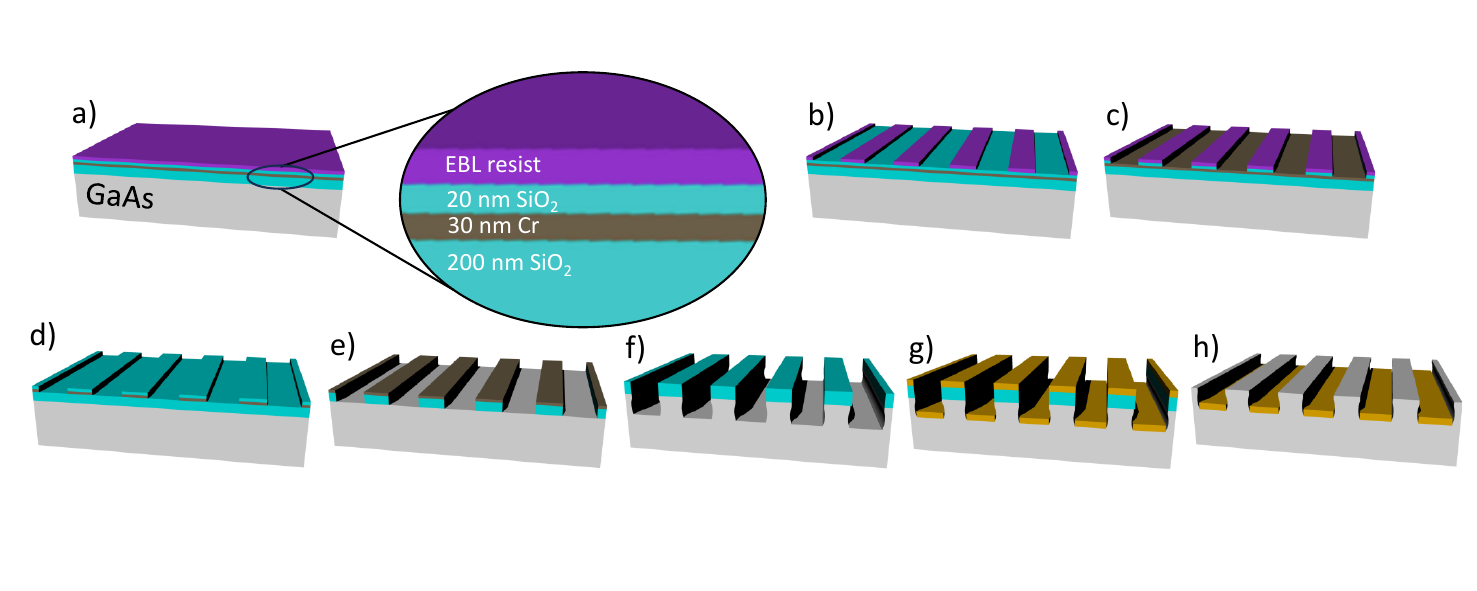}
\caption{\label{fig:S1}Flow chart of metalMHCG fabrication steps: a) double side polished GaAs wafer covered with set of mask layers and electron beam sensitive resist; b) EBL fabricated pattern; c) pattern transfer to upper SiO$_2$ layer; d) pattern transfer to Cr layer; e) pattern transfer to lower SiO$_2$ layer; f) pattern transfer to GaAs substrate; g) Au e-beam evaporation and SiO$_2$ mask lift-off; h) fabricated GaAs-Au metalMHCG structure.}
\end{figure*}

The metalMHCG structure was fabricated using a combination of plasma enhanced chemical vapour deposition (PECVD), electron beam lithography (EBL), inductively coupled plasma-reactive ion etching (ICP-RIE), and e-beam physical vapor deposition (EBPVD), following the technological steps illustrated in Fig. \ref{fig:S1}. First, the top of a double-sided polished GaAs substrate was covered with a triple layer consisting of SiO$_2$/Cr/SiO$_2$, to act as an etch mask. The thicknesses of the three layers of SiO$_2$/Cr/SiO$_2$ were \SI{200}{\nano \meter}, \SI{30}{\nano \meter}, and \SI{20}{\nano \meter}, respectively (Fig. \ref{fig:S1}a). The prepared substrate was then spin-coated with a 200-nm thick layer of AR-P 6200.9 e-beam resist, followed by EBL patterning and development in AR 600-546 for \SI{60}{\second} at 22$^{\circ}$C (Fig. \ref{fig:S1}b). The EBL-fabricated pattern, representing the desired MHCG design, was then transferred sequentially by plasma etching to a 20-nm thick SiO$_2$ layer (Fig. \ref{fig:S1}c), a 30-nm thick Cr layer (Fig. S1d), and finally to a 200-nm thick SiO$_2$ layer (Fig. \ref{fig:S1}e) that served as the main GaAs etch mask. The SiO$_2$ layers were etched using CHF$_3$/CF$_4$ chemistry at a temperature of 20$^{\circ}$C, \SI{100}{\watt} RF power, and a chamber pressure of mTorr \SI{20}{\mt}. The Cr layer was etched using Cl$_2$/O$_2$ chemistry at 20$^\circ$C, with a PICP of \SI{800}{\watt}, PRF of \SI{50}{\watt}, and chamber pressure of \SI{10}{\mt}. The upper layer of SiO$_2$ (\SI{20}{\nano \meter} thick) and the layer of Cr were sacrificial layers that allowed the EBL-fabricated pattern to be transferred to the lower SiO$_2$ layer (\SI{200}{\nano \meter} thick) etch mask. The GaAs substrate was then plasma etched (Fig. \ref{fig:S1}f) using BCl$_3$/Ar chemistry at a temperature of 10$^\circ$C, with a PICP of \SI{180}{\watt}, PRF of \SI{100}{\watt}, and a chamber pressure of \SI{5}{\mt}, resulting in slightly concave profiles of the etched side walls. The SiO$_2$ hard mask remaining unetched. The substrate was then diced into 5×5mm samples, and selected samples were e-beam evaporated with gold (Fig. \ref{fig:S1}g), with thicknesses ranging from \SI{50}{\nano \meter} to \SI{250}{\nano \meter} in \SI{50}{\nano \meter} increments. The fabrication process was completed by removing the SiO$_2$ hard mask (Fig. \ref{fig:S1}h) and covering  the metalMHCG structure with a layer of Au,  deposited on top using buffered hydrofluoric acid (BOE) solution. 

The first generation of metalMHCGs had the expected concave cross-section of the side walls, similar to those presented in Fig. 4a, spatial parameters ($L, F, H$) close to the designed parameters. However, the non-rectangular cross-section of the semiconductor stripes resulted in a shift of the transmittance maximum relative to the expected wavelength, leading to a maximum transmittance of polarized light of only 80\%. To optimize the transmittance through the metalMHCG, we used scanning electron microscopy (SEM) images to accurately capture the cross-section of the semiconductor and metal stripes. 

\section{Transmittance measurements}\label{sec:s2}

\begin{figure*}[h!]
\centering
\includegraphics[width=1.0\textwidth]{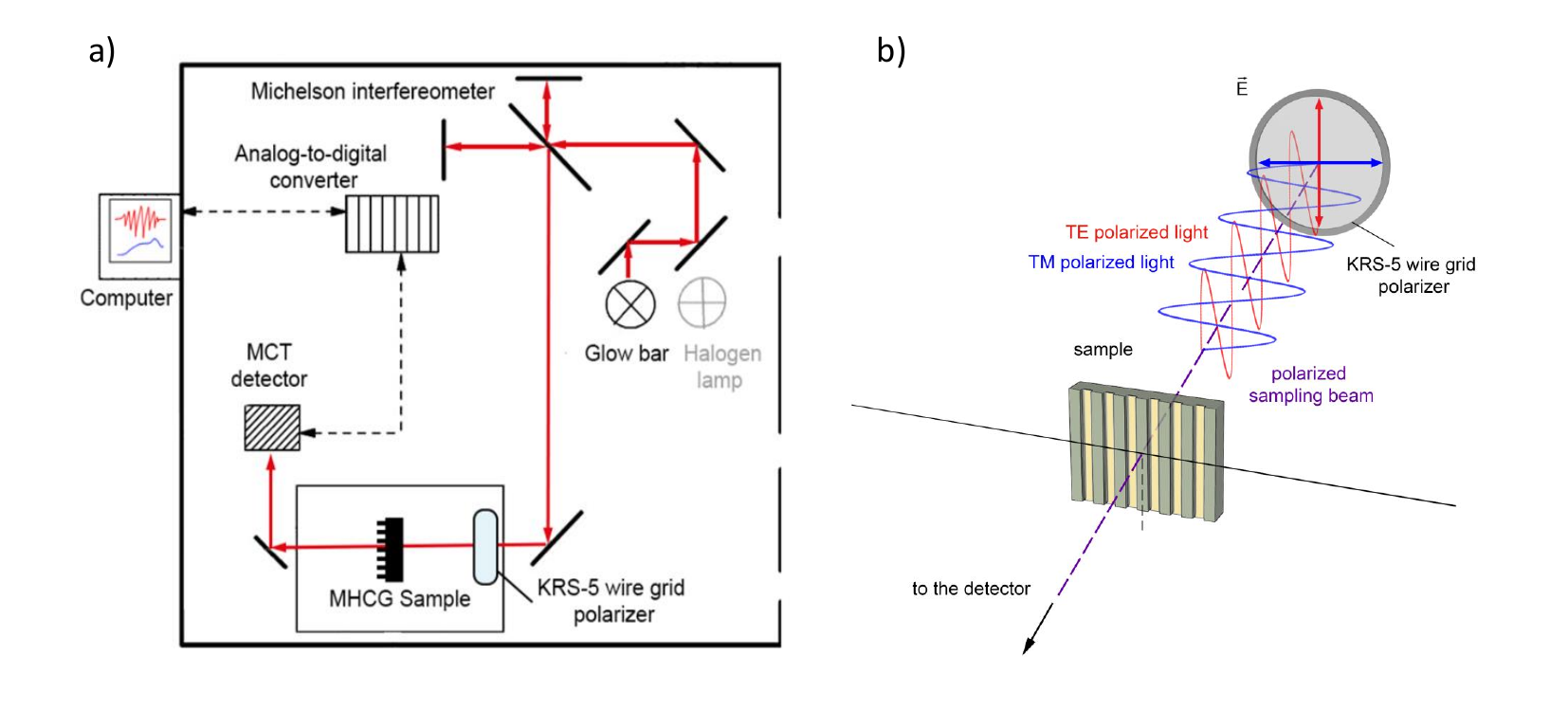}
\caption{\label{fig:S2}a) Schematic view of the beam path in a transmittance experiment with FTIR setup; b) close-up of polarizer mounting showing the direction of stripes in the sample.}
\end{figure*}

\begin{figure*}[h!]
\centering
\includegraphics[width=1.0\textwidth]{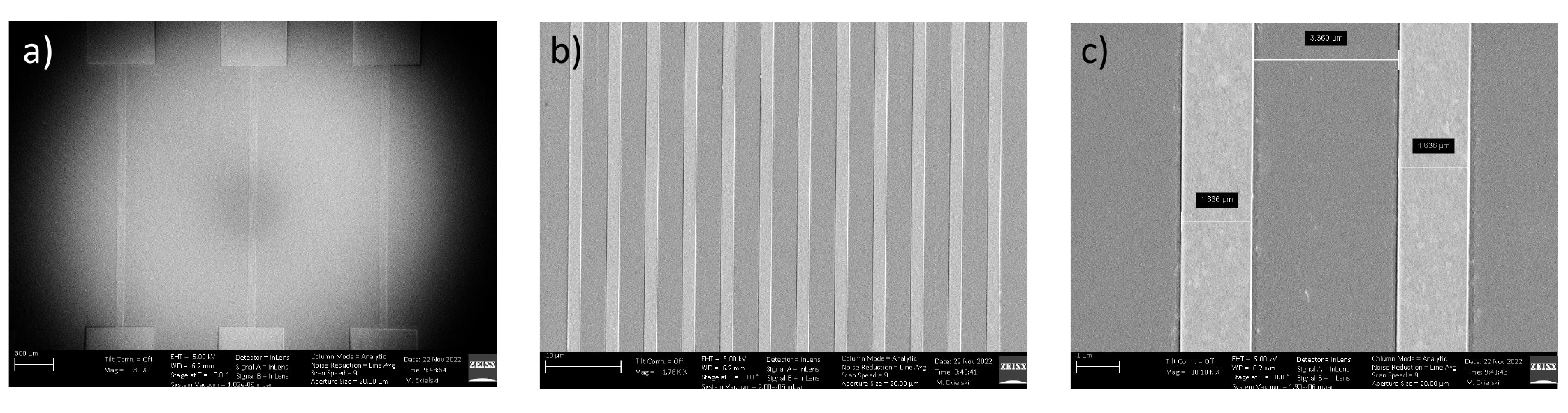}
\caption{\label{fig:S3}Top view scanning electron microscope (SEM) images at various degrees of magnification. Images show structures with 50-nm thick metal stripes prepared for electrical characterisation.}
\end{figure*}

\begin{figure*}[h!]
\centering
\includegraphics[width=0.5\textwidth]{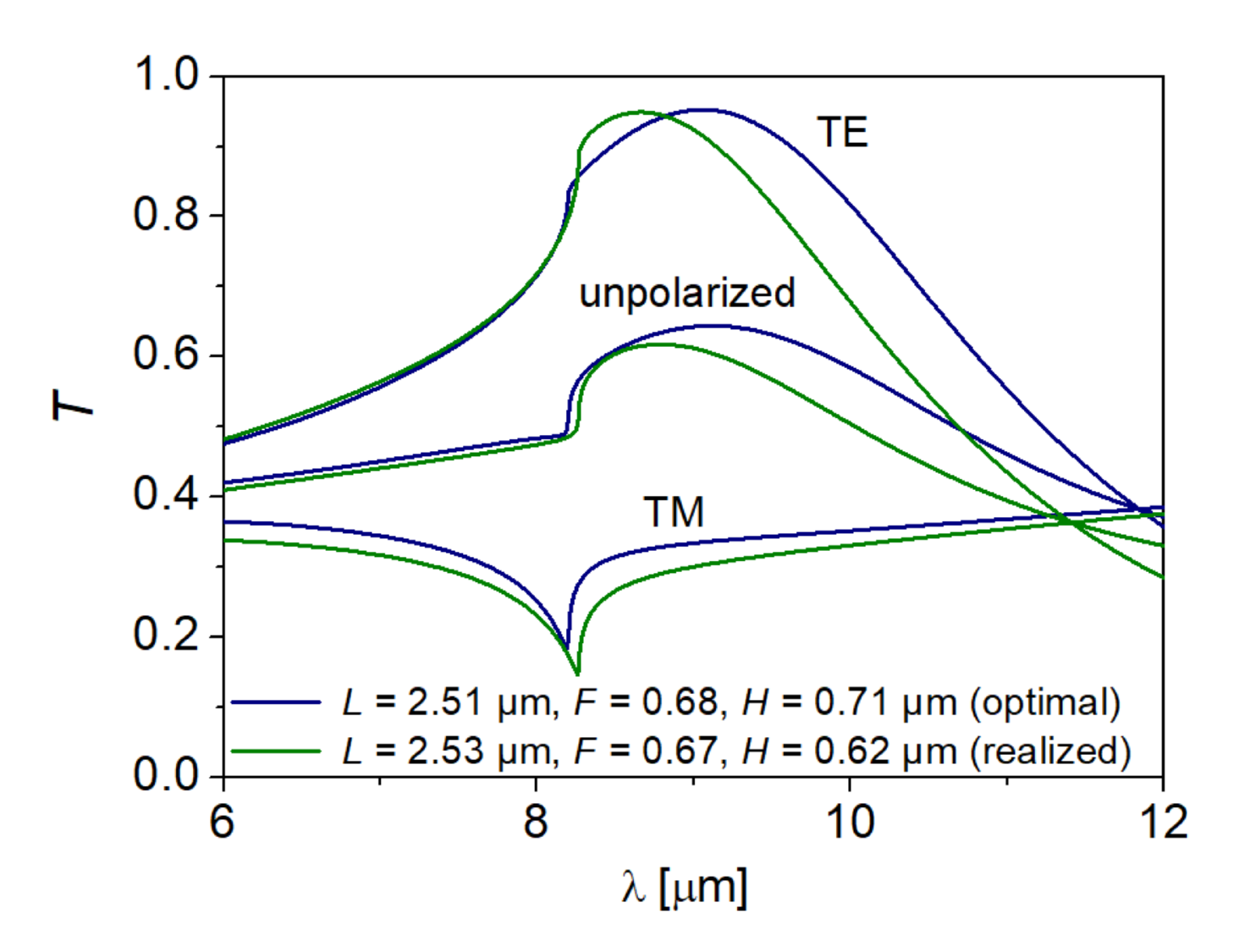}
\caption{\label{fig:OptvsRe}Calculated transmittance spectra of metalMHCG in the case of TE-polarized (TE), TM-polarized (TM) and unpolarized light incidence. The spectra are calculated for real-world cross-sections of semiconductor and metal stripes with the geometrical parameters indicated in the figure and $H_\textup{m} = \SI{200}{\nano \meter}$. Blue represents optimal configuration and green the configuration realised in the experiment.}
\end{figure*}

\begin{figure*}[h!]
\centering
\includegraphics[width=0.5\textwidth]{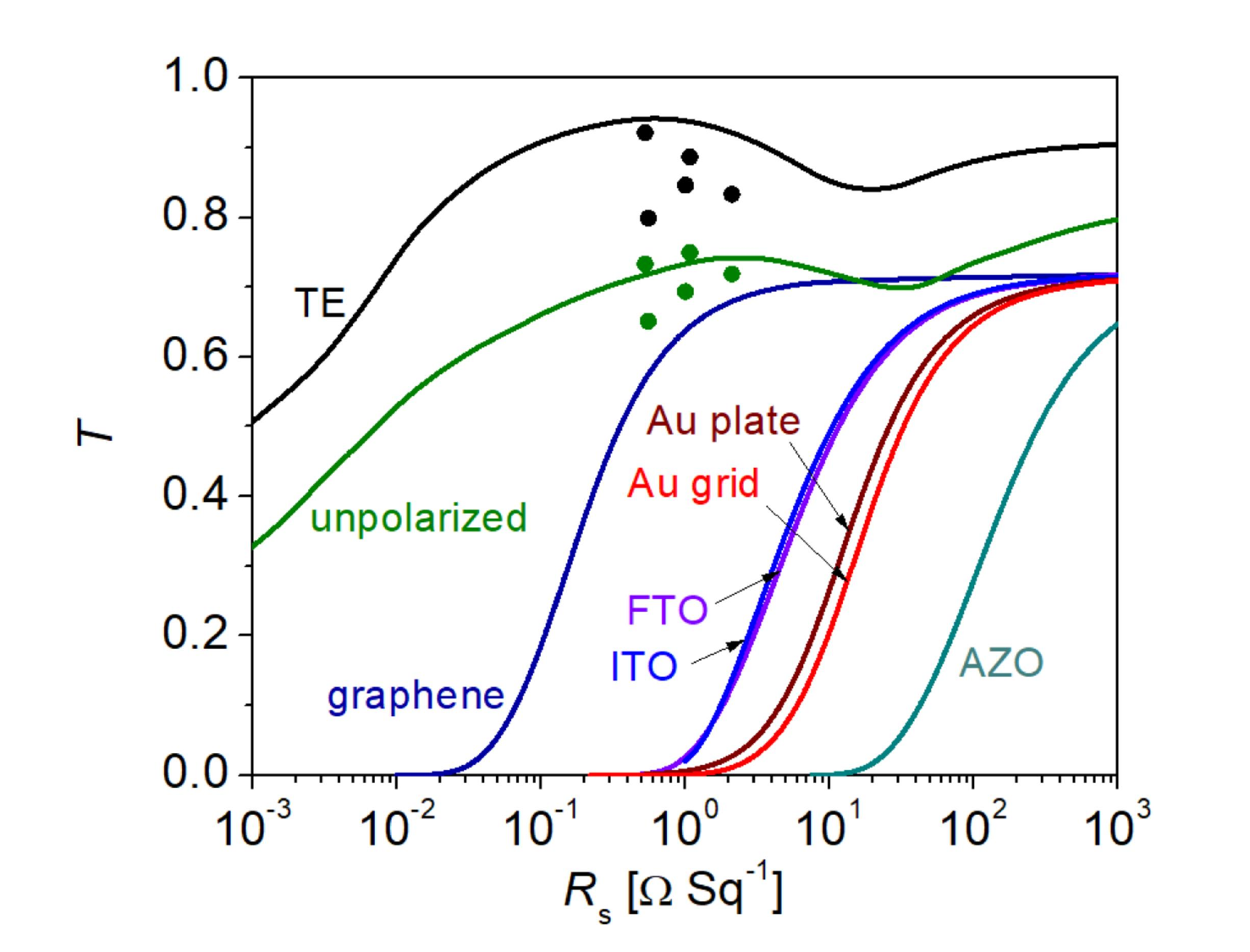}
\caption{\label{fig:S4} Measured absolute transmittance versus sheet resistance of metalMHCGs measured in this work for TE polarized (black dots) and unpolarized light (green dots). Black and green lines  represent the calculated dependences of absolute transmittance for the GaAs-based metalMHCG versus sheet resistance for TE transmittance and unpolarized transmittance, respectively. Other lines represent calculated absolute transmittance for graphene, metal oxide, gold plate, and gold grids all deposited on GaAs wafer versus sheet resistance for their refractive indices corresponding to the infrared spectral range and lowest reported resistivities \protect\citeS{Wang_Overvig_Shrestha_Zhang_Wang_Yu_Dal_Negro_2017, Farhan_Zalnezhad_Bushroa_Sarhan_2013, Khalilzadeh-Rezaie_Oladeji_Cleary_Nader_Nath_Rezadad_Peale_2015, Lalasari_Arini_Andriyah_Firdiyono_Yuwono_2018, Shkondin_Takayama_Panah_Liu_Larsen_Mar_Jensen_Lavrinenko_2017,Shen_Zhang_Lu_Jiang_Yang_2010, Ullah_Nawi_Witjaksono_Tansu_Khattak_Junaid_Siddiqui_Magsi_2020, Liu_Deshmukh_Salowitz_Zhao_Sobolev_2022}}
\end{figure*}

Transmittance measurements were conducted using a Vertex 80v vacuum Fourier Transform Infrared spectrometer (FTIR) from Bruker. Due to the many limitations and difficulties of working in the mid-infrared spectral region, it was necessary to use a Fourier spectrometer instead of a simpler monochromator-based setup. The advantages of using the FTIR approach over the dispersive setup have been extensively studied and described \protect\citeS{Thermofisher, Griffiths_A._2007}. Figure \ref{fig:S2}a shows the experimental setup for transmittance measurements. The sample light generated by a polychromatic source (either a halogen or a glow bar, depending on the spectral range) was guided by parabolic golden mirrors to the Michelson interferometer and then focused onto the sample at a normal incident angle, creating a roughly 1-mm diameter spot  entirely contained within the 5×\SI{5}{\milli \meter} metalMHCG sample. The sample was placed on a mounting holder with a slit slightly larger than the diameter of the focused spot, which allowed light to transmit through the entire sample and exit from the etched grating. The transmitted light was then directed to a HgCdTe (MCT) liquid-nitrogen cooled detector and transferred as an electrical response into an analog-to-digital converter. The signal was later modified using a 3-Term Blackman-Harris apodization function or an adequate zero-filling factor and converted into a spectrum using Fast Fourier Transform. To perform polarization-resolved measurements, a KRS-5 wire-grid polarizer was inserted into the beam path before the light reached the sample, as shown in Fig. 1b. Using this experimental procedure, we were able to obtain spectra in the 1.3-\SI{16}{\micro \meter} spectral range with a \SI{4}{\per \centi \meter} resolution \protect\citeS{Motyka_Sek_Janiak_Misiewicz_Klos_Piotrowski_2011, Motyka_Misiewicz_2010}. 

\bibliographystyleS{unsrt}
\bibliographyS{Supplement}
%\bibliography{Supplement}

\end{document}